# Nitrogen-Doped Ti$_3$C$_2$T$_x$ Coated with a Molecularly Imprinted Polymer as Efficient Cathode Material for Lithium-Sulfur Batteries

*Feng Yan,[1],[†] Liqiang Lu,[2],[†] Chongan Ye,[1] Qi Chen,[1] Sumit Kumar,[3] Wenjian Li,[2] Hamoon Hemmatpour,[1] Konstantinos Spyrou,[4] Sytze de Graaf,[1] Marc C. A. Stuart,[5] Bart J. Kooi,[1] Dimitrios P. Gournis,[4] Katja Loos,[1] Yutao Pei,[2]\* and Petra Rudolf[1]\**

[1]Zernike Institute for Advanced Materials, University of Groningen, Nijenborgh 4, 9747AG Groningen, the Netherlands

[2]Engineering and Technology Institute Groningen, University of Groningen, Nijenborgh 4, 9747AG Groningen, the Netherlands

[3]Electrical Engineering Division, Department of Engineering, University of Cambridge, Cambridge, CB21PZ, UK

[4]Department of Materials Science and Engineering, University of Ioannina, 45110 Ioannina, Greece

[5]Groningen Biomolecular Sciences and Biotechnology Institute, University of Groningen, Nijenborgh 4, 9747AG Groningen, the Netherlands

---

[†] These authors contributed equally to this work.

**ABSTRACT**

Due to their high energy density (2600 Wh kg$^{-1}$), low cost, and low environmental impact, lithium-sulfur batteries are considered a promising alternative to lithium-ion batteries. However, their commercial viability remains a formidable scientific challenge mainly because of the sluggish reaction kinetics at the cathode and the so-called "shuttling effect" of soluble polysulfides, which results in capacity decay and a shortened lifespan. Herein, molecular imprinting with Li$_2$S$_8$ as a target molecule in combination with a two-dimensional material, MXene, is proposed to overcome these issues. Molecularly imprinted polymer-coated nitrogen-doped Ti-based MXene was successfully synthesized and demonstrated to exhibit an appealing electrochemical performance, namely a high specific capacity of 1095 mAh g$^{-1}$ at 0.1 C and an extended cycling stability (300 mAh g$^{-1}$ at 1.0 C after 300 cycles). X-ray photoelectron spectroscopy was applied to elucidate the underlying mechanisms and proved that Li$_2$S$_8$-imprinted polymer polyacrylamide serves as a polysulfide trap through strong chemical affinity towards the long-chain lithium polysulfides, while N-doped Ti-based MXene promotes the redox kinetics by accelerating the conversion of lithium polysulfides. This distinct interfacial strategy is expected to result in more effective and stable Li-S batteries.

**INTRODUCTION**

Rechargeable batteries with long-term cycling stability, high energy density and low cost have been considered as key solution to overcome the ever-increasing energy storage demands for portable electronic devices.[1] Lithium-sulfur (Li-S) batteries are particularly attractive because of their high theoretical specific capacity and energy density (2567 Wh kg$^{-1}$), non-toxicity, and the natural abundance of sulfur.[2,3] However, Li-S batteries have so far been hampered in their progress towards commercialization by several fundamental issues. On one hand, the insulating nature of sulfur and of the solid-state discharge products ($Li_2S_2$ and $Li_2S$) cause sluggish electrode kinetics, resulting in large electrode polarization and poor rate capacity.[4] On the other hand, lithium polysulfide chains (LiPS) ($Li_2S_x$, 2<x<8) form at the cathode, shuttle through the electrolyte, and react at the anode, causing severe deterioration of the battery.[5] Therefore, immobilizing the LiPS and boosting the redox kinetics of sulfur species are the key to advancing Li-S batteries.[6]

Recent extensive efforts have mainly focused on developing conductive matrices as effective sulfur host to accelerate electron transfer or as electro-catalyst to improve the redox kinetics.[7] In this context a new class of two-dimensional compounds, MXenes, has emerged as promising cathode material for Li-S batteries because of their high conductivity in the core structure and abundant polar functional groups on the surface.[8,9] The general formula of MXenes is $M_{n+1}X_nT_x$, where M stands for transition metals such as Ti or V, X represents C or N, and the bond between them mainly contributes to the conductivity, while T denotes functional polar groups at the surface, such as -OH, O, Cl, and F, that provide strong chemisorption sites for lithium polysulfides.[10] However, once inserted in the polymer matrix the MXene nanosheets tend to restack due to the hydrogen bonding and van der Waals interaction between the polar groups on the surface and this restacking seriously hinders the infiltration of the electrolyte between the nanosheets and limits the full utilization of the functional surfaces, resulting in poor electrochemical performance.[11] Several strategies have been applied to control the structure and enhance the accessible surface of $Ti_3C_2T_x$ nanosheets, including manufacturing 0D nanodots[12], 1D nanoribbons[9] and 3D porous frameworks[8]. Overall, the polar terminal groups on the surface of the 2D MXene nanosheets promote the formation of channels, which facilitate the diffusion of $Li^+$ in the electrode, limit the "shuttle effect" of polysulfides, and increase the sulfur loading, while restricting volumetric expansion.[13] In addition to building a 3D porous structure, modification of the surface functional groups and heteroatom doping of the nanosheets with elements like N and S can further enhance the catalytic effect by accelerating the conversion of lithium polysulfide and lower the $Li^+$ diffusion barrier.[14,15]

Molecularly imprinted polymers (MIPs), created by reacting the monomers in the presence of a template molecule, are able to selectively recognize that template molecule.[16] Because of this predetermined selectivity and outstanding efficiency, MIPs have been extensively explored for a wide range of applications, such as drug delivery,[17] catalysis,[18] sensors,[19] and environmental remediation.[20] However, the strategy of exploiting molecularly imprinted polymers as cathode materials to trap the long-chain lithium polysulfides in order to suppress the shuttle effect has been barely researched so far.[21] Yan *et al.*[22] have reported for the first time that carbon nanotubes coated with a molecularly imprinted polymer performed outstandingly well in the capture of long chain polysulfides, demonstrating the potential of MIPs for application in Li-S batteries.[23]

In this work, we combine a MIP with porous N-doped $Ti_3C_2T_x$ (N-$Ti_3C_2T_x$) to fabricate a highly promising cathode material. We started with pyrolysis of melamine intercalated $Ti_3C_2T_x$ nanosheets, then polymerized acrylamide with $Li_2S_8$ as template molecule and subsequently removed the template molecules to obtain vacant binding sites. The resulting $N-Ti_3C_2T_x$ coated with the imprinted polymer shows an appealing electrochemical performance; the underlying mechanism was elucidated with the help of X-ray photoelectron spectroscopy.

## Results and Discussion

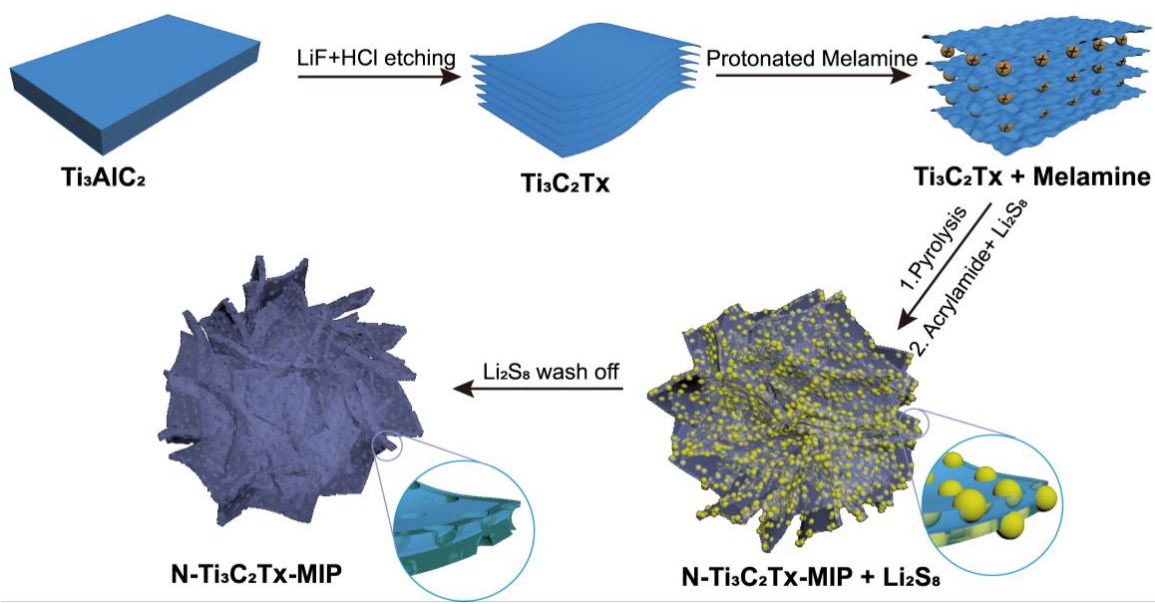

**Scheme 1. Schematic illustration of the preparation of a composite consisting of nitrogen-doped MXene coated with the molecularly imprinted polymer ($N-Ti_3C_2T_x$-MIP)**

Molecularly imprinted polymer with $Li_2S_8$ recognition characteristics were obtained by polymerization of acrylamide monomers with $Li_2S_8$ as the target molecular template on the surface of crumpled N-doped $Ti_3C_2T_x$; the resulting composite will be called $N-Ti_3C_2T_x$-MIP in the following. Scheme 1 provides a schematic description of the synthetic procedure, and the detailed protocol is included in the Supporting Information. The morphology of $N-Ti_3C_2T_x$-MIP was firstly investigated by SEM and TEM. The SEM image in Figure 1(a) shows that $N-Ti_3C_2T_x$-

MIP has a porous structure with a highly exfoliated and crumpled appearance. The larger magnification SEM image in Figure 1(b) shows pores with a diameter of ~ 3 µm. The origin of this texture lies presumably in the violent expansion upon annealing combined with the removal of the polar groups (-F, -OH, -Cl), which weakened the interaction between the layers and thereby reduced the probability of restacking. The pristine $Ti_3C_2T_x$ nanosheets, for which the SEM image are presented in Figure S1 (a) and (b) exhibit instead the typical restacked morphology with compact platelets.[24] An additional difference in the structure of N-$Ti_3C_2T_x$-MIP compared to the etched MAX phase are the macropores between the layers, which seemingly stem from the freeze-drying step after the molecularly imprinted polymer coating was applied. The high-angle annular dark-field scanning transmission electron microscopy (HAADF-STEM) image in Figure 1(c) shows the edge of N-$Ti_3C_2T_x$-MIP together with the selected area electron diffraction (SAED) pattern (Figure 1(d)). One distinguishes an around 7.0 nm thick amorphous layer at the edge, which confirms that N-$Ti_3C_2T_x$ is indeed coated with a polymer. The in-plane area exhibits a polycrystalline structure and the SAED pattern has the hexagonal symmetry of $Ti_3C_2T_x$.[14]

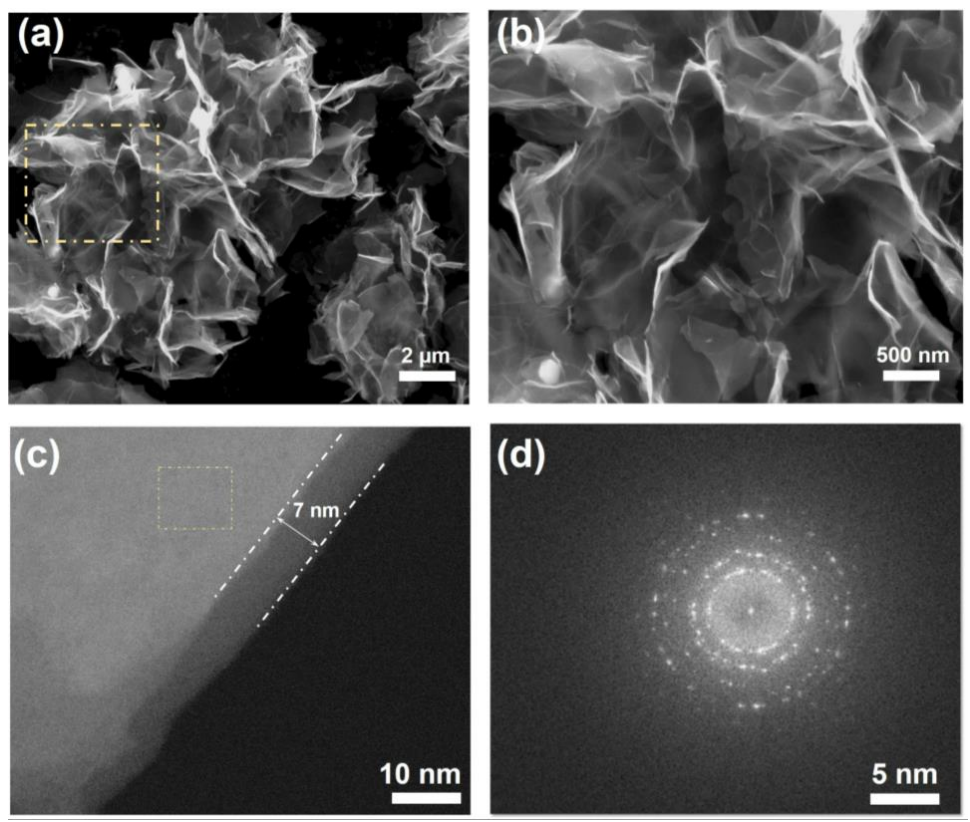

**Figure 1. Morphology and structure of the molecularly imprinted polymer-coated N-doped Ti$_3$C$_2$T$_x$: (a) and (b) SEM images; (c) HAADF-STEM image; (d) corresponding selected area electron diffraction (SAED) pattern collected from the area outlined in yellow in the HAADF-STEM image (c).**

After the melting diffusion of sulfur in N-Ti$_3$C$_2$T$_x$-MIP that yields the sample labeled N-Ti$_3$C$_2$T$_x$-MIP/S, SEM and TEM were performed to check whether sulfur was uniformly distributed. The SEM image in Figure 2(a) shows no significant sulfur particle aggregation in the crumpled structure, suggesting that a homogeneous sulfur loading was achieved. Furthermore, the HRTEM image of N-Ti$_3$C$_2$T$_x$-MIP/S in Figure 2(b) shows sulfur particles with the diameter around 5 nm on the surface of the N-Ti$_3$C$_2$T$_x$-MIP sheets.[25] The different contrast in the micrograph in Figure 2(c) allows to recognize three layers, namely the 3 nm thick sulfur layer located at the outside of the 7 nm thick MIP layer, which in turn coats the N-Ti$_3$C$_2$T$_x$ sheet, all in good agreement with the results from Figure 1(c).

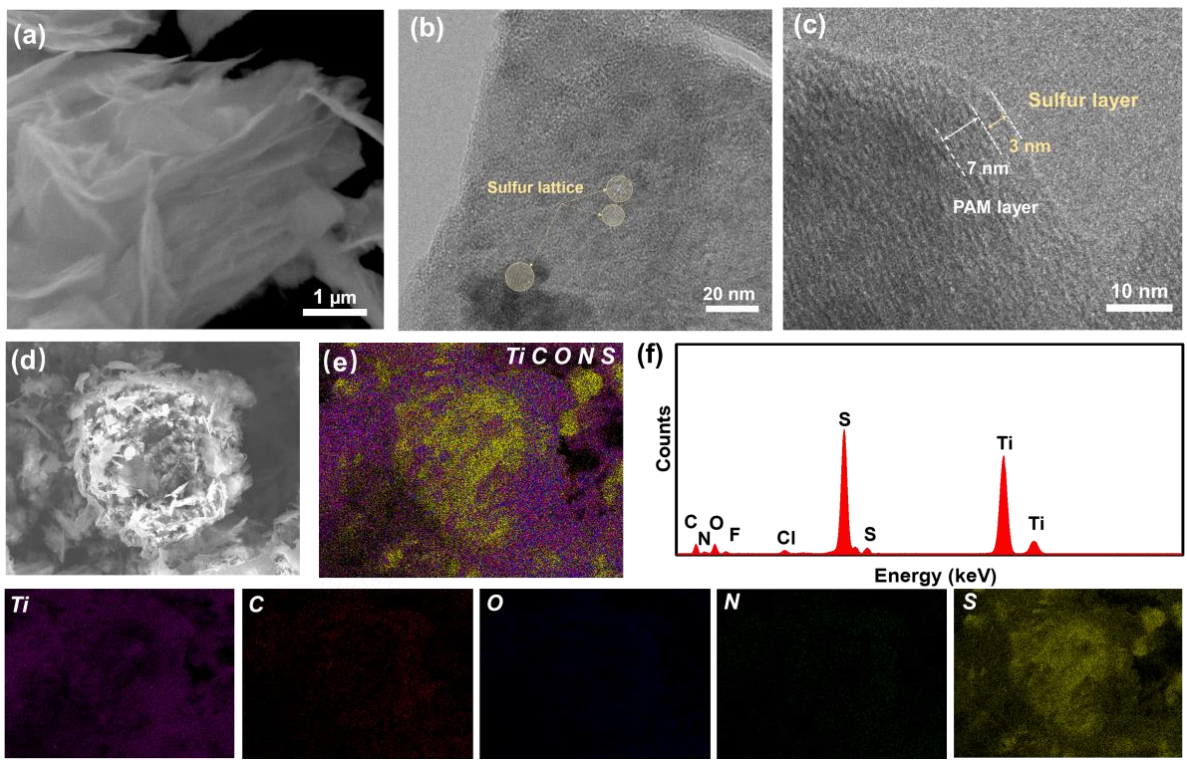

**Figure 2. Morphology of the molecularly imprinted polymer-coated N-doped Ti$_3$C$_2$T$_x$ and elemental mapping after sulfur loading: (a) SEM image; (b) and (c) HRTEM images; (d) SEM image and (e) and corresponding superimposed elemental mapping; (f) SEM-EDS spectrum collected on (d) Bottom row: EDS mapping of the Ti, C, O, N and S elemental distribution collected on (d).**

In order to verify the distribution of the various elements in N-Ti$_3$C$_2$T$_x$-MIP/S, SEM-EDS elemental mapping was performed. A typical EDS spectrum of the area marked with a dashed yellow line in the SEM image in Figure 2(d) is presented in Figure 2(f). The elemental maps of Ti, C, O, N and S, imaged separately in Figure 2, bottom row, show that all five elements are evenly distributed. In addition, the superposition of the elemental maps, displayed in Figure 2(e), demonstrates that the sulfur particles are homogeneously trapped in the porous N-$Ti_3C_2T_x$-MIP matrix.

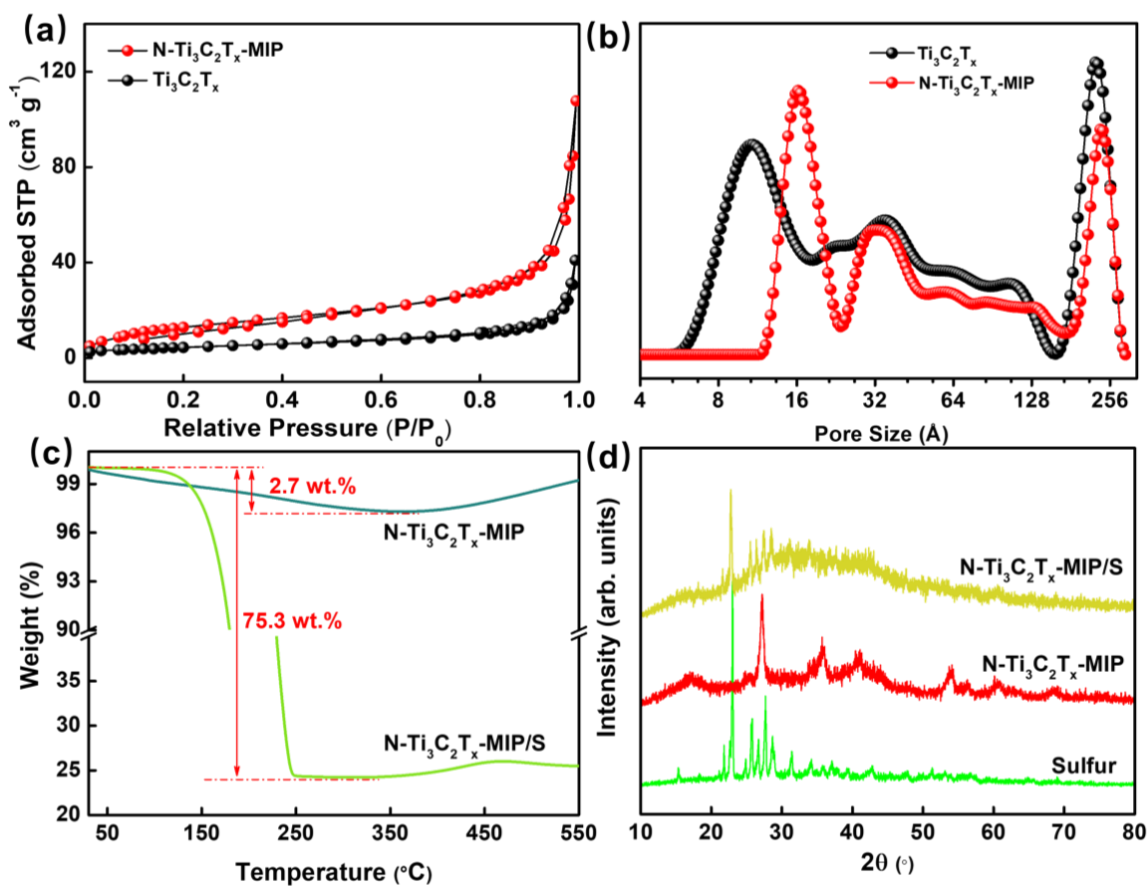

**Figure 3.** (a) $N_2$ adsorption-desorption isotherms of pristine $Ti_3C_2T_x$ and of molecularly imprinted polymer-coated N-doped $Ti_3C_2T_x$; (b) pore size distribution as deduced from non-local density functional theory (NLDFT); (c) TGA curves of molecularly imprinted polymer-coated N-doped $Ti_3C_2T_x$ before and after sulfur loading; (d) X-ray diffraction patterns of a sulfur crystal, and of molecularly imprinted polymer-coated N-doped $Ti_3C_2T_x$ before and after sulfur loading.

In order to determine how the specific surface area of N-Ti$_3$C$_2$T$_x$-MIP differs from that of pristine Ti$_3$C$_2$T$_x$ and to retrieve information on the pore size distribution, N$_2$ adsorption-desorption measurements were performed; the corresponding isotherms are plotted in Figure 3(a). A type II isotherm can be observed with an important hysteresis in the range of 0.5-1.0 of relative pressure for both Ti$_3$C$_2$T$_x$ and N-Ti$_3$C$_2$T$_x$-MIP. The type H3 hysteresis loop, commonly ascribed to the presence of mesoporous and macroporous structure is much wider for N-Ti$_3$C$_2$T$_x$-MIP than for Ti$_3$C$_2$T$_x$, and the slope of the N-Ti$_3$C$_2$T$_x$-MIP isotherm in the higher relative pressure region dramatically steeper than that of the starting material, implying an about 10-fold increase in mesoporous and macroporous pore volume from 0.093 cm$^3$ g$^{-1}$ in Ti$_3$C$_2$T$_x$ to 0.923 cm$^3$ g$^{-1}$ in N- Ti$_3$C$_2$T$_x$-MIP. The specific surface area, which was found to amount to 32.6 m$^2$ g$^{-1}$ for Ti$_3$C$_2$T$_x$, increased to 137.8 m$^2$ g$^{-1}$ after modification. These data indicate that N$_2$ can only adsorb on the external surface of Ti$_3$C$_2$T$_x$ and not in the interlayer space because of restacking, while the opening of the interlayer space by polymer coating allows for N$_2$ penetration. The pore size distribution in Ti$_3$C$_2$T$_x$ and in N-Ti$_3$C$_2$T$_x$-MIP was determined with non-local density functional theory (NLDFT), and the results are shown in the inset of Figure 3(b). In the micropore region, a broad peak centered at 1.1 nm can be observed for Ti$_3$C$_2$T$_x$, while for N-Ti$_3$C$_2$T$_x$-MIP a much narrower peak centered at 1.6 nm can be seen, in accordance with the size of Li$_2$S$_8$. These results indicate that the defects on the surface of pristine Ti$_3$C$_2$T$_x$ have been covered with molecularly imprinted polymer with vacant recognition binding sites, whose size coincides with the template molecule.

To estimate the amount of sulfur loaded in N-Ti$_3$C$_2$T$_x$-MIP/S, TGA was performed on N-Ti$_3$C$_2$T$_x$-MIP and N-Ti$_3$C$_2$T$_x$-MIP/S, and the results are presented in Figure 3(c). For N-Ti$_3$C$_2$T$_x$-MIP, the TGA curve shows a gradual weight loss below 350 °C, totaling 2.7 wt.%;

this loss mirrors the decomposition of the $Li_2S_8$-imprinted polyacrylamide on the surface. For N-$Ti_3C_2T_x$-MIP/S, the TGA curve shows a large weight loss of 75.3 wt.% of the total mass in the range 200-350 °C; this loss originates from the evaporation of sulfur loaded in the composite and indicates that the sulfur stored in the N-$Ti_3C_2T_x$-MIP/S amounts to 72.6 wt.%. The slight increase of weight at higher temperature for both samples attracted our attention, as it has been reported also by others,[26] and could be due to impurities either in the $N_2$ gas or introduced during the synthesis procedure, since partial degradation of (N-)$Ti_3C_2T_x$ to $TiO_2$ occurs (*vide infra*).

The crystal structure of $Ti_3C_2T_x$ and its derivatives was further investigated by XRD, as shown in Figure S2 and Figure 3(d). The diffraction peaks of commercial $Ti_3AlC_2$, in Figure S2, match well with those of a MAX phase material (JCPDS, no. 52-7874), while after etching with HCl/LiF, the characteristic peaks of $Ti_3C_2T_x$ are clearly observed, in agreement with a previous report,[27] indicating that delaminated $Ti_3C_2T_x$ was successfully synthesized. In addition, the emergence of a peak at 27.2°, ascribed to the $TiO_2$,[28] demonstrates that the $Ti_3C_2T_x$ matrix was partially oxidized during the synthetic protocol for the modification, as confirmed also by the TEM results in Figure S3, in which the lattice spacing is 0.35 nm, corresponding to the (101) planes of $TiO_2$.[29]

In order to verify which chemical species are present in the near-surface region of the composites, X-ray photoelectron spectroscopy was employed. The overview spectra of $Ti_3C_2T_x$, N-$Ti_3C_2T_x$-MIP, and N-$Ti_3C_2T_x$-MIP/S, shown in Figure S4, attest to the presence of all the expected elements. In fact, in addition to the spectral signatures of Ti, C, O, Cl, and F, present in $Ti_3C_2T_x$, a nitrogen peak is seen for N-$Ti_3C_2T_x$-MIP and for N-$Ti_3C_2T_x$-MIP/S also a S signal is clearly visible, indicating the successful introduction of nitrogen (N doping and N-containing polymer coating) and sulfur loading for the respective compounds.[14]

Since the introduction of nitrogen in the MXene is of great importance for improving the electrochemical behavior of the electrode, the detailed XPS spectrum of the N1$s$ core level region of N-Ti$_3$C$_2$T$_x$ and N-Ti$_3$C$_2$T$_x$-MIP was collected, and the results are shown in Figure 4(a). For N-Ti$_3$C$_2$T$_x$, four contributions are needed to achieve a satisfactory fit: the spectral signature of the Ti-N bond centered at a BE of 396.9 eV confirms the successful doping of N into the Ti$_3$C$_2$T$_x$ matrix.[30] The contributions at BEs of 398.4, 399.9, and 401.4 eV are respectively ascribed to the pyridinic N[30], pyrrolic N,[31] and graphite N,[32]; the elemental content of nitrogen in N-Ti$_3$C$_2$T$_x$ was 1.2 at.%, as shown in Table S1. After coating with molecularly imprinted polymer, the N1$s$ core level spectrum of N-Ti$_3$C$_2$T$_x$-MIP shows a significant increase in spectral intensity at 400.0 eV, ascribed to O=C-NH$_2$ moieties, demonstrates the successfully coating of polyacrylamide onto the surface of N-Ti$_3$C$_2$T$_x$ sheets, and the elemental content of nitrogen increased to 8.4 at.%.

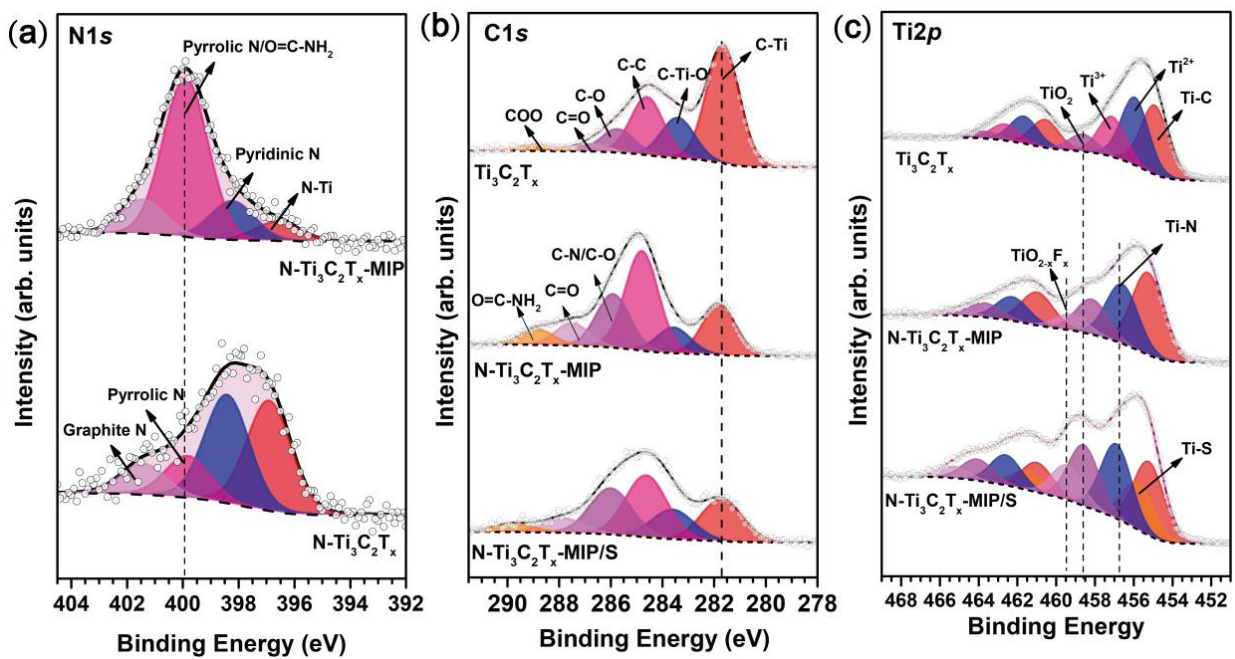

**Figure 4. X-ray photoelectron spectra of (a) the N1$s$ core level region of N-doped Ti$_3$C$_2$T$_x$ before and after coating with the molecularly imprinted polymer; (b) C1$s$ and (c) Ti2$p$ core-level regions of pristine Ti$_3$C$_2$T$_x$, and of N-doped Ti$_3$C$_2$T$_x$ coated with the molecularly imprinted polymer before and after sulfur loading.**

In order to track the changes in chemical bonding and the consequent changes in interaction introduced at each synthesis step, the XPS spectra of the C1$s$ and Ti2$p$ core level regions of pristine Ti$_3$C$_2$T$_x$, N-Ti$_3$C$_2$T$_x$-MIP and N-Ti$_3$C$_2$T$_x$-MIP/S were further explored, and the respective spectra are presented in Figure 4(b) and 4(c). Six types of carbon bonds contribute to the C1$s$ core level photoemission line of pristine Ti$_3$C$_2$T$_x$, as shown in Figure 4(b). These are C-C, C-O, and C=O and C(O)O bonds that give rise to the spectral intensity in the range of 284.7-288.5 eV and stem from acid etching. At BEs of 281.7 and 283.5 eV, there are the signatures of C-Ti and C-Ti-O bonds that constitute the core structure of Ti$_3$C$_2$T$_x$; these two components make up 62.6 % of the total C1$s$ intensity.[35] In the Ti2$p$ core level spectrum of pristine Ti$_3$C$_2$T$_x$ (Figure 4(c)), the four doublets with 2$p_{3/2}$ peaks located at 454.9, 455.9, 457.1, and 458.6 eV, can be assigned to Ti-C, Ti$^{2+}$, Ti$^{3+}$, and TiO$_2$ (Ti$^{4+}$) bonds, respectively, consistent with a previous report.[36]

As for N-Ti$_3$C$_2$T$_x$-MIP, the XPS spectrum of the C1$s$ core level region shows a significant decrease in intensity of the components due to C-Ti and C-Ti-O bonds, while the spectral intensity in the range of 284.7-289.1 eV increases, as expected for an efficacious coating of polyacrylamide on the surface of the nitrogen-doped Ti$_3$C$_2$T$_x$. The appearance of a component due to Ti-N bonds in the Ti2$p$ core level spectrum of N-Ti$_3$C$_2$T$_x$-MIP testifies to the successful insertion of nitrogen in the Ti$_3$C$_2$T$_x$ matrix. In addition, one observes that the Ti2$p_{3/2}$ peak of N-Ti$_3$C$_2$T$_x$-MIP is shifted towards higher binding energy as compared to that of pristine Ti$_3$C$_2$T$_x$; this shift can be attributed to the doping effect of nitrogen on the Ti$_3$C$_2$T$_x$ matrix and the removal of terminal groups. Nitrogen in the matrix creates a more electron deficient environment at the nearby Ti atoms since it acts as electron withdrawing atom given its high electronegativity (3.0). The removal of terminal of -OH groups also results in more electron deficient Ti.[37,11]

After sulfur loading, the detailed XPS spectrum of the S2$p$ core level region of N-Ti$_3$C$_2$T$_x$-MIP/S (Figure S5) shows four doublets: the ones peaked at BEs of 163.7 and 165.2 eV correspond[14] respectively to S-S/C-S and C-S-O bonds, while the contribution peaked at a BE of 162.1 eV is due to S-Ti bonds[38] and implies that sulfur passes through the polyacrylamide layer and forms bonds with the N-doped Ti$_3$C$_2$T$_x$ matrix. At the higher BEs the peak at 168.5 eV is signature of oxidized sulfur species (SO$_2$/SO$_3$).[39] The shape of C1$s$ core level line of N-Ti$_3$C$_2$T$_x$-MIP/S (Figure 4(b)) confirms the presence of C-S and C-Ti-S bonds mirrored by the intensity increase of the components at 283.8 and 285.9 eV.[40] As seen in Figure 4(c), a Ti-S component appears at a BE of 456.6 eV in the Ti2$p$ core level spectrum of N-Ti$_3$C$_2$T$_x$-MIP/S, reconfirming[41] that sulfur binds to the metal atoms after penetrating through the porous MIP layer. One also observes that the intensity of the TiO$_2$ component located at 458.5 eV[42] increases and that a new peak arises at a BE of 459.6 eV; the latter is ascribed[43] to TiO$_{2-x}$F$_x$. These spectral features point to partial oxidation during the heat treatments for N-doping and sulfur loading, in agreement with the XRD and TEM data.

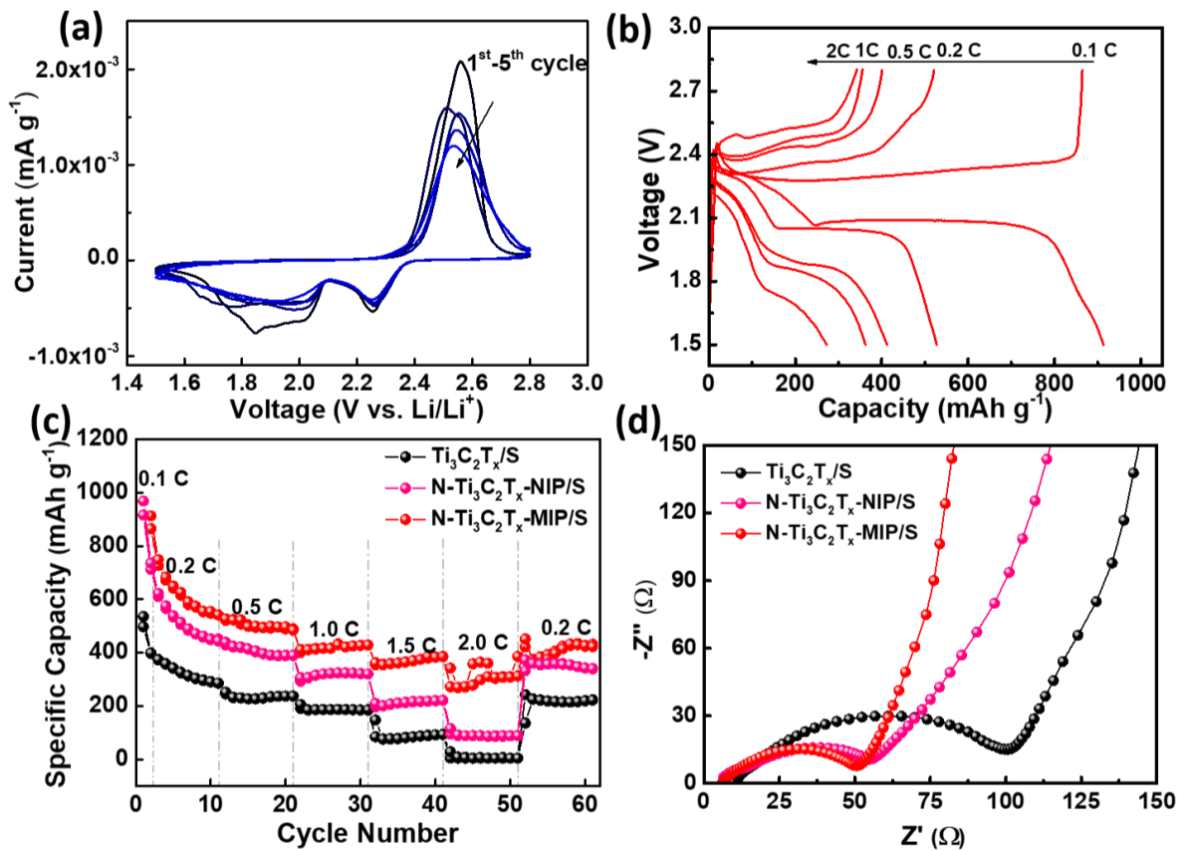

**Figure 5.** Electrochemical measurements: (a) CV profiles of the N-Ti$_3$C$_2$T$_x$-MIP/S cathode at a scan rate of 0.1 mV s$^{-1}$ in a potential window from 1.7 to 2.8 V; (b) galvanostatic charge-discharge profile of the N-Ti$_3$C$_2$T$_x$-MIP/S electrodes cycled between 1.7 and 2.8 V while increasing the current rate; (c) rate capacities of Li-S cells with, Ti$_3$C$_2$T$_x$/S, N-Ti$_3$C$_2$T$_x$-NIP/S and N-Ti$_3$C$_2$T$_x$-MIP/S cathodes; (d) electrochemical impedance spectra of cells with Ti$_3$C$_2$T$_x$/S, N-Ti$_3$C$_2$T$_x$-NIP/S and N-Ti$_3$C$_2$T$_x$-MIP/S cathodes at the stage of discharging to 1.7 V.

As has been reported by Natu *et al.*,[33] nitrogen doping of MXene can induce a high catalytic reactivity for the sulfur redox reaction because it lowers the dissociation barrier for lithium polysulfide. Therefore, we expect that the N-containing polymer (MIP) binds lithium polysulfide through recognition by the specific binding sites created by the template, while N-doping of Ti$_3$C$_2$T$_x$ can effectively improve the redox reaction with its superior catalytic reactivity. Thus the combination of both functions should improve the electrochemical performance of the Li-S battery.[34] Therefore, having demonstrated the success of the synthesis of this novel electrode material, we went on to prove its effectiveness when integrated in coin-type Li-S battery cells with the N-Ti$_3$C$_2$T$_x$-MIP/S as cathode and lithium metal anodes, by investigating the electrochemical performance; the results are presented in Figure 5. The electrochemical stability of N-Ti$_3$C$_2$T$_x$-MIP was first evaluated by cyclic voltammetry (CV), and the five initial cycles as shown in Figure 5(a). In the first and second cycles, the CV profiles of the N-Ti$_3$C$_2$T$_x$-MIP/S present three reduction peaks during discharging and one dominant oxidation peak during charging. The three reduction peaks, located at around 2.27, 2.05 and 1.84 V, are attributed respectively to the conversion of cyclo-S$_8$ to long-chain lithium polysulfides (Li$_2$S$_n$, $4 \leq n \leq 8$), the further reduction of long-chain lithium polysulfides to L$_2$S$_2$, and the solid-state conversion from L$_2$S$_2$ to Li$_2$S. In addition, slight polarization in the first activation cycle is observed, which is due to the strong adsorption ability of the N-Ti$_3$C$_2$T$_x$-MIP matrix, while the positive shift in the reduction peaks and the negative shift in the oxidation peak indicate an improved polysulfide redox kinetics. Furthermore, the CV curves of the cells collected during the third, fourth and fifth cycle overlap well, confirming that the sulfur cathodes exhibit excellent electrochemical reversibility and stability.

The galvanostatic charge/discharge behavior of the cells with N-Ti$_3$C$_2$T$_x$-MIP/S electrode was evaluated at different current densities (from 0.1 to 2.0 C), as shown in Figure 5(b), and exhibits two typical plateaus at 2.30 and 2.05 V, in a good agreement with the CV profile. Although the discharge plateaus are gradually decreasing with increasing the current density, they are still clearly distinguished and rather stable even at the high rates of 0.5 C to 2.0 C, indicating fast reaction kinetics of the N-Ti$_3$C$_2$T$_x$-MIP/S composites at high current densities.[44]

In order to be able to distinguish the effects of the presence of the polymer alone from those caused by the molecular imprinting, we decided to compare cells with N-Ti$_3$C$_2$T$_x$-MIP/S to those with electrodes with uncoated, undoped, sulfur-loaded MXene, Ti$_3$C$_2$T$_x$/S, as well as to cells with electrodes, where the coating of N-doped Ti$_3$C$_2$T$_x$ with polyacrylamide was performed without the Li$_2$S$_8$ template molecule before proceeding with the S loading (labeled as N-Ti$_3$C$_2$T$_x$-NIP/S). The rate performances of the N-Ti$_3$C$_2$T$_x$-MIP/S, N-Ti$_3$C$_2$T$_x$-NIP/S, and Ti$_3$C$_2$T$_x$/S electrodes at different current rates are shown in Figure 5(c). The N-Ti$_3$C$_2$T$_x$-MIP/S cathode delivers an initial discharge capacity of 1095 mAh g$^{-1}$ at 0.1 C, pointing to a high utilization of sulfur. The discharge capacity faded gradually at 0.1 C, but when further increasing the current density to 0.2 C, 0.5 C, and 2.0 C, the discharge capacities delivered were respectively 518, 423, and 268 mAh g$^{-1}$, and exhibited good stabilities. When the current rate was returned to 0.2 C, the electrode recovered a discharge capacity of 415 mAh g$^{-1}$, without showing signs of degradation. The high-rate performance is in agreement with the voltage profiles in Figure 5(b). Of the three cathodes, the electrode with N-Ti$_3$C$_2$T$_x$-MIP/S achieved the highest specific capacity; this can be rationalized by the good contact between sulfur and N-Ti$_3$C$_2$T$_x$-MIP as well as to the ease with which the electrolyte penetrates the electrode material through the molecularly imprinted polymer and is involved in the redox reaction on the nitrogen-doped Ti$_3$C$_2$T$_x$, as anticipated. Our hypothesis is supported by the fact that for the N-Ti$_3$C$_2$T$_x$-NIP/S electrode, where the surface is also covered with polyacrylamide but without any specifically designed binding sites, the performance is not as good, implying that it is more difficult for electrolyte and lithium polysulfides to penetrate the polymer. The poorer rate performance of the Ti$_3$C$_2$T$_x$/S electrode when compared to the N-Ti$_3$C$_2$T$_x$-MIP/S one can be ascribed to the restacking of Ti$_3$C$_2$T$_x$ flakes, which prevents the electrolyte from penetrating, limits the effective surface available for the redox reaction, and lowers the surface area of contact with sulfur.

Additional insight into the improved electrochemical performance of the N-Ti$_3$C$_2$T$_x$-MIP/S electrode can be gained from the electrochemical impedance spectroscopy (EIS) analysis of the N-Ti$_3$C$_2$T$_x$-MIP/S, N-Ti$_3$C$_2$T$_x$-NIP/S and Ti$_3$C$_2$T$_x$/S cathodes, shown in Figure 5(d). The impedance plots show the typical semicircle at high frequency, while at low frequency the impedance increases linearly. The semicircle represents the interfacial charge-transfer impedance, whereas the linear slope at low frequency reflects the Li ion diffusion in the cathodes.[45] The impedance plot of N-Ti$_3$C$_2$T$_x$-MIP/S shows a semicircle, which is slightly smaller than that of N-Ti$_3$C$_2$T$_x$-NIP/S, while both are significantly smaller than that of Ti$_3$C$_2$T$_x$/S, meaning that the N-Ti$_3$C$_2$T$_x$-MIP/S has slightly lower interfacial charge-transfer resistance than that of N-Ti$_3$C$_2$T$_x$-NIP/S electrodes, again pointing to faster redox reaction kinetics on N-Ti$_3$C$_2$T$_x$-MIP/S. Since N-Ti$_3$C$_2$T$_x$-MIP/S and N-Ti$_3$C$_2$T$_x$-NIP/S have higher surface area than Ti$_3$C$_2$T$_x$/S, the better contact between sulfur and the polymer-covered MXenes accelerates the charge transfer and results in a much smaller resistance. In addition, nitrogen doping of MXene favors electron transport. The steep increase of the impedance at low frequency implies that the Li$^+$ diffusion resistance of the N-Ti$_3$C$_2$T$_x$-MIP/S electrode is much smaller than those of the N-Ti$_3$C$_2$T$_x$-NIP/S and Ti$_3$C$_2$T$_x$/S electrodes, another demonstration of the superior electrochemical performance of N-Ti$_3$C$_2$T$_x$-MIP/S. The fast reaction kinetics and Li$^+$ diffusion of N-Ti$_3$C$_2$T$_x$-MIP/S are in accordance with the rate performances.

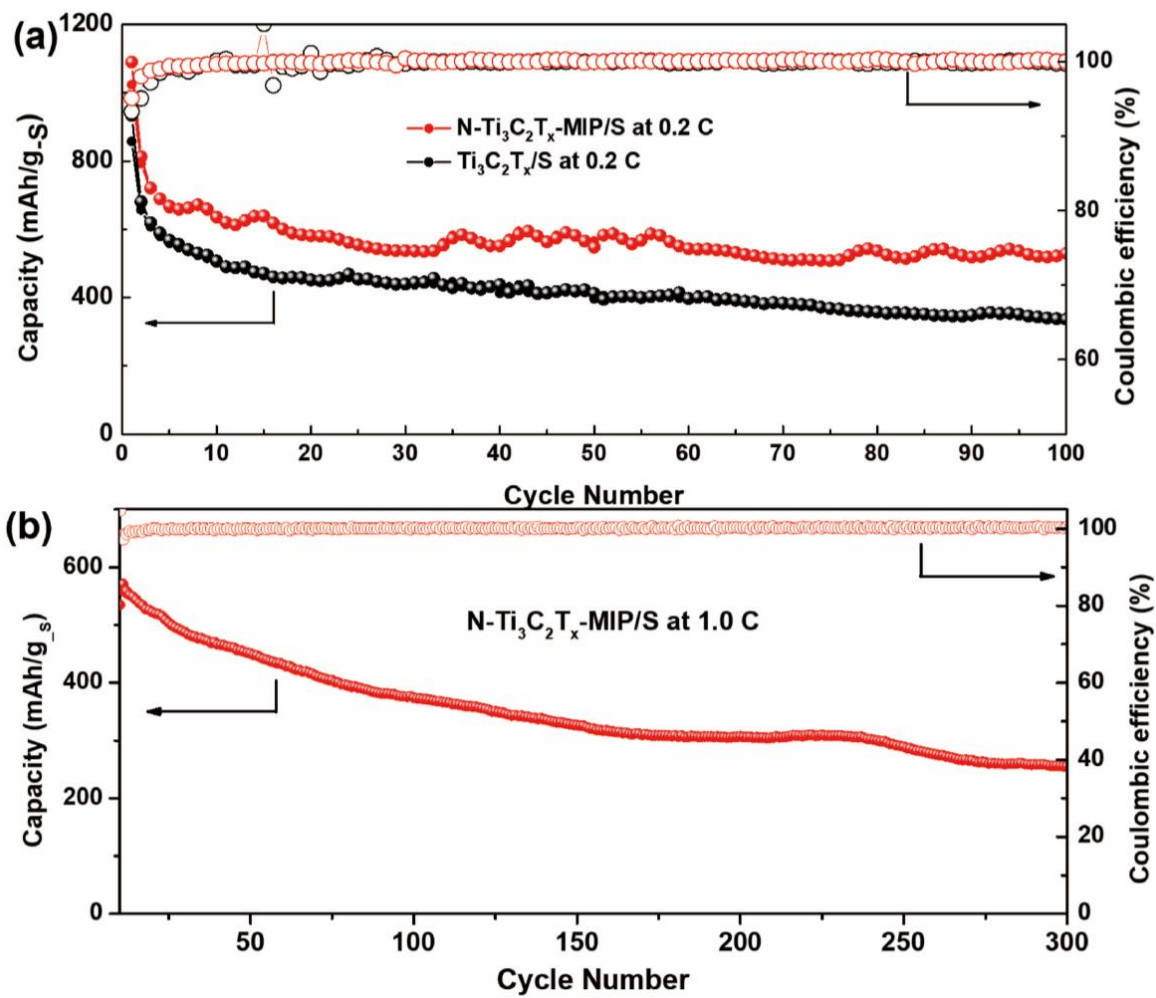

**Figure 6.** **(a) Discharge/charge capacities and Coulombic efficiency of cathodes consisting of sulfur-loaded $Ti_3C_2T_x$ and of sulfur-loaded N-doped $Ti_3C_2T_x$ coated with molecularly imprinted polymer measured at rates of 0.2 C. (b) Cyclic stability of Li-S battery cathode assembled with N-$Ti_3C_2T_x$-MIP/S measured at 1.0 C.**

Another aspect that is important in the evaluation of the design of the new N-$Ti_3C_2T_x$-MIP electrode material is its reversible cycling performance at both low and high current rates. We compared the electrochemical performances of N-$Ti_3C_2T_x$-MIP and $Ti_3C_2T_x$/S electrodes at 0.2 C for 100 cycles; the results are shown in the Figure 6(a). The average Coulombic efficiencies of the sulfur cathodes are almost 100 %, indicating that the charge and discharge processes take place quickly and effectively in both electrochemical measurements. The $Ti_3C_2T_x$/S electrode achieved a discharge capacity of 676 mAh g$^{-1}$ in the first cycle but only 343 mAh g$^{-1}$ after 100 cycles. In contrast, the N-$Ti_3C_2T_x$-MIP/S electrode exhibited an initial discharge capacity of 681 mAh g$^{-1}$, which decreased to 525 mAh g$^{-1}$ over 100 cycles. This improved cycling performance encouraged us to check the N-$Ti_3C_2T_x$-MIP/S electrode's cycling stability at 1.0 C, as shown in Figure 6(b). The N-$Ti_3C_2T_x$-MIP/S electrode's discharge capacity, initially 571 mAh g$^{-1}$, decreased to 300 mAh g$^{-1}$ over 300 cycles, which corresponds to a decay rate of 0.16 % per cycle. This improved performance of the N-$Ti_3C_2T_x$-MIP/S electrode demonstrates that our design approach that combines higher conductivity, higher porosity, and more efficient lithium polysulfide trapping is going in the right direction.

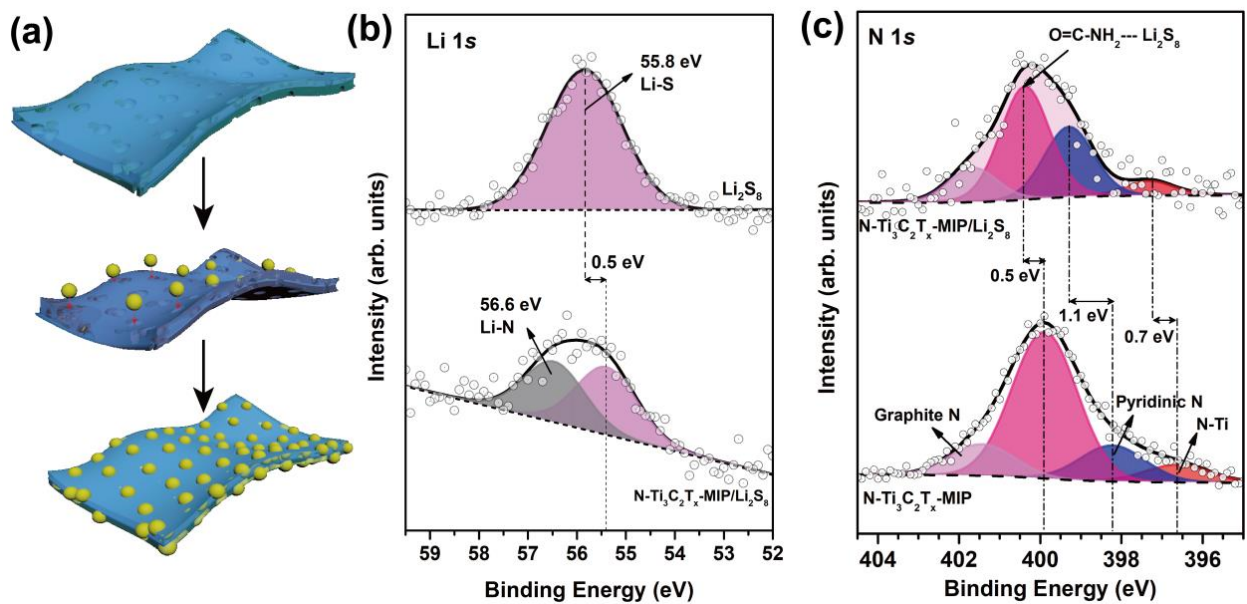

**Figure 7. (a) Schematic illustration of the $Li_2S_8$ recognition and capture process with the molecularly imprinted polymer composite N-$Ti_3C_2T_x$-MIP; X-ray photoelectron spectra of**

**(b) the Li1*s* core level region of Li$_2$S$_8$ and of N-Ti$_3$C$_2$T$_x$-MIP/Li$_2$S$_8$; (c) the N1*s* core-level region of N-Ti$_3$C$_2$T$_x$-MIP and of N-Ti$_3$C$_2$T$_x$-MIP/Li$_2$S$_8$.**

To further investigate the unique chemisorptive property of the N-Ti$_3$C$_2$T$_x$-MIP nanosheets and their interaction with lithium polysulfide, X-ray photoelectron spectroscopy was employed to gain insight into the interactions between N-Ti$_3$C$_2$T$_x$-MIP and Li$_2$S$_8$ in this molecular recognition and adsorption process illustrated schematically in Figure 7(a). The XPS spectra of the Li1*s* core level region of the Li$_2$S$_8$ molecule and N-Ti$_3$C$_2$T$_x$-MIP/Li$_2$S$_8$ are presented in Figure 7(b); the Li1*s* spectrum of Li$_2$S$_8$ shows a symmetric peak at a BE of 55.8 eV, attributed to the Li-S bond. This same component is seen to have shifted to 55.3 eV in the spectrum of N-Ti$_3$C$_2$T$_x$-MIP/Li$_2$S$_8$, which also features a new component at a BE of 56.6 eV, due to Li-N bonds.[46] The shift of the Li-S component implies that Li$^+$ in the lithium polysulfide interacts with the N-containing groups of the MIP coating layer or the matrix, in other words when Li$^+$ chemisorbs on N-Ti$_3$C$_2$T$_x$-MIP, it withdraws electronic charge from N-containing groups, thus the Li atom becomes more electron-rich, and the peak shifts to the lower binding energy.[37] The appearance of Li-N indicates the chemisorption of lithium polysulfides on N-Ti$_3$C$_2$T$_x$-MIP, which helps reduce the shuttle effect of the battery. A redistribution of charge in the N environment due to the interaction between Li$_2$S$_8$ and N-Ti$_3$C$_2$T$_x$-MIP matrix is clearly visible in the XPS spectrum of the N1*s* core level region shown in Figure 7(c). After Li$_2$S$_8$ exposure, the components due to O=C-NH$_2$, pyrodinic N and N-Ti bonds are shifted to higher binding energies by respectively 0.5 eV, 1.1 and 0.7 eV. The shifts of the latter two components are ascribed to the N-doping in the Ti$_3$C$_2$T$_x$ increases the electron density of the matrix, and thereby causes a stronger interaction with Li atoms from Li$_2$S$_8$.[47] The shift of the first component is instead due to the interaction of Li$_2$S$_8$ and the O=C-NH$_2$ groups located on the wall of imprinted polymer pore.[48] Together the results from the electrochemical tests and the XPS results demonstrate that N-doping in the $Ti_3C_2T_x$ matrix and applying a molecularly imprinted polymer coating on the surface anchor lithium polysulfide effectively, results in an improved electrochemical performance of Li-S battery. Therefore, it can be concluded that introducing MIP in the cathode material of Li-S battery is a winning strategy.

**Conclusions**

In summary, a novel Li-S cathode material consisting of N-doped $Ti_3C_2T_x$ coated with polyacrylamide, formed in the presence of $Li_2S_8$ to create specific binding sites for lithium polysulfides, was successfully synthesized. Comprehensive physical characterization confirmed that this synthesis strategy gives rise to a porous structure with a highly exfoliated and crumpled appearance. The high specific surface area and specific binding sites for $Li_2S_8$ make this material an outstanding candidate as a host scaffold for sulfur loading in Li-S batteries. Li-S batteries with N-$Ti_3C_2T_x$-MIP/S composite as electrode delivered a high specific capacity of 1095 mAh $g^{-1}$ at 0.1 C and extended cycling stability (300 mA h $g^{-1}$, at 1.0 C rate after 300 cycles). The improved electrochemical performance can be ascribed to the suppression of the shuttle effect by specific binding sites for lithium polysulfides of the $Li_2S_8$-imprinted polymer, as well as to the enhanced catalytic reactivity imparted to the $Ti_3C_2T_x$ matrix by nitrogen doping, which effectively improves the redox reaction. Furthermore, X-ray photoelectron spectroscopy corroborates the nitrogen doping and proves the interaction between $Li_2S_8$ and the electrode material, supporting that the N-doped in the $Ti_3C_2T_x$ matrix and the N-containing molecularly imprinted polymer both play important role in improving the electrochemical properties. Therefore, this novel strategy for improving electrode performance is anticipated to push the rational design of electrode materials and opens the road toward better Li-S batteries.

**Methods**

**Materials.** MAX ($Ti_3AlC_2$) phase in powder form ($\leq$ 40 μm) was acquired from Y-Carbon Ltd., Ukraine. Lithium fluoride (LiF, ~300 mesh), hydrochloric acid (HCl, 37.0 %), ethanol ($\geq$ 99.0 %), melamine ($\geq$ 99.0 %), lithium sulfide ($Li_2S$, 99.9 %), sulfur (S, powder, 99.5 %), potassium persulfate ($K_2S_2O_8$, 99.9 %), acrylamide (AA, 99.0 %), anhydrous n,n-dimethylformamide (DMF, 99.9 %), anhydrous n-methyl-2-pyrrolidone (NMP, 99.5 %) and anhydrous carbon disulfide ($CS_2$, 99.7 %) were purchased from Sigma-Aldrich. Carbon black was acquired from Fisher Scientific. All the chemicals were used as received. Milli-Q water (resistivity 18 MΩ·cm, 25 °C) was freshly produced before use.

**Synthesis of $Ti_3C_2T_x$ flakes.** $Ti_3C_2T_x$ was synthesized by etching of Al atomic layers from the MAX ($Ti_3AlC_2$) phase. For this, 8.0 g of LiF was added slowly to 100 mL HCl (9.0 M) while magnetically stirring for 30 min, then 5.0 g of $Ti_3AlC_2$ powder was gradually added to the solution to avoid overheating given that the reaction is exothermic. **(Caution! Highly corrosive acid, HF, is formed in the reaction. Contact with the skin must be avoided by wearing HF acid resistant gloves, and this etching process should be performed only in a well-ventilated fume hood.)** The mixture was stirred at 350 rpm for 24 h at 40 °C. After reaction, the mixture was repeatedly washed with Milli-Q water, and centrifuged at 4500 rpm for 10 min until a deep dark supernatant and a swollen sediment were observed. The sediment was redispersed in 200 mL of Milli-Q water and sonicated in an ice bath with Ar flow for 2 h. In order to eliminate the non-etched MAX phase from the multilayer $Ti_3C_2T_x$ particles, the mixture solution was centrifuged at 3500 rpm for 1 h, and the turbid upper black liquid was collected. By washing several times with Milli-Q water the pH of the supernatant was brought to 5.5, whereafter the Ti$_3$C$_2$T$_x$ flakes were collected by centrifugation at 12000 rpm and vacuum dried at room temperature overnight. The dry black powder was stored in a vacuum desiccator for further use.[49]

**Synthesis of the N-Ti$_3$C$_2$T$_x$-MIP composite**. The protocol for preparing the composite of N-doped Ti$_3$C$_2$T$_x$ and molecularly imprinted polyacrylamide, labeled N-Ti$_3$C$_2$T$_x$-MIP in this work, is illustrated in Scheme 1. 3D porous nitrogen-doped Ti$_3$C$_2$T$_x$ was synthesized by intercalating positively charged melamine between the Ti$_3$C$_2$T$_x$ layers, and then calcinating the composite at 550 °C in Ar atmosphere. The synthesis procedure comprised three steps: (1) Positively charged melamine was prepared. For this 1.0 g of melamine was dispersed in 15.0 mL of ethanol under sonication and magnetic stirring for 1 h, then 1.5 mL of hydrochloric acid dropwise to the solution and stirred for another 1 h; the mixture was transferred to a petri dish to let the solvent evaporate in the oven 70 °C overnight. The dry, positively charged melamine solid was ground into a powder and washed several times with ethanol. (2) The treated melamine was intercalated between Ti$_3$C$_2$T$_x$ layers. For this, 25 mg of Ti$_3$C$_2$T$_x$ powder was dispersed into 50 mL Milli-Q water and 100 mg positively charged melamine was dissolved in 50 mL 0.1 M dilute hydrochloric acid; the two liquids were mixed, an operation that turned Ti$_3$C$_2$T$_x$ into a floating flocculent, and stirred for 2 h. Then the Ti$_3$C$_2$T$_x$ derivative was isolated by centrifugation and freeze-dried. (3) The intercalated Ti$_3$C$_2$T$_x$ was calcinated by inserting 100 mg into a chemical vapor deposition (CVD) setup (Carbolite Gero Ltd.), heating it in Ar atmosphere from 25 to 550 °C, at a rate of 5 °C min$^{-1}$, and keeping it at 550 °C for 4 h.

Then N-Ti$_3$C$_2$T$_x$-MIP was then synthesized by polymerizing acrylamide with Li$_2$S$_8$ as the target template molecule on the surface of the porous nitrogen-doped Ti$_3$C$_2$T$_x$. Firstly, Li$_2$S$_8$ was synthesized in the Ar-filled glove box by adding 5.61 g (0.175 M) of sulfur and 1.15 g (0.025 M) of Li$_2$S (molar ratio 7:1) to 50 mL of anhydrous DMF and magnetically stirring at 50 °C until the powder was fully dissolved. Secondly, 20 mg of N-doped $Ti_3C_2T_x$ was vigorously dispersed in 10 mL anhydrous DMF, then 1.0 mM acrylamide and 2.5 mM $Li_2S_8$ (5 mL) were added to the dispersion. The polymerization of acrylamide was initiated by introducing 0.03 mM $K_2S_2O_8$ at 70 °C in Ar atmosphere. After polymerization for 12 h, $Li_2S_8$ was removed from the MIP by repeated washing with DMF and centrifugation until the filtrate turned from yellow to colorless. For a control experiment, the non-imprinted polymer coated nitrogen-doped $Ti_3C_2T_x$ was synthesized by following the same procedure without introducing $Li_2S_8$; this sample is denoted as $N-Ti_3C_2T_x$-NIP in this work.

**Charging the composite of N-Ti₃C₂Tₓ-MIP with sulfur.** The composite of N-doped $Ti_3C_2T_x$ and molecularly imprinted polyacrylamide was charged with sulfur as follows: 150 mg of N-$Ti_3C_2T_x$-MIP was first uniformly dispersed in 4 mL of ethanol under stirring for 0.5 h, and then 7 mL of carbon disulfide ($CS_2$) were added to the mixture. After stirring for 10 min, 360 mg of sulfur was added to the mixture, followed by stirring for 1 h. $CS_2$ and ethanol were then evaporated at 35 °C under stirring, whereafter a black powder of N-$Ti_3C_2T_x$-MIP/S was obtained. In order to favor a more homogeneous sulfur distribution, the as-obtained black powder was heated at 155 °C for 24 h in Ar atmosphere.[50] N-$Ti_3C_2T_x$-NIP and $Ti_3C_2T_x$ were charged with sulfur in the same fashion.

**Testing for the electrochemical performance.** Electrodes for lithium-sulfur batteries were synthesized by mixing 80 wt.% N-$Ti_3C_2T_x$-MIP/S with 10 wt.% of carbon black, and 10 wt.% of polyvinylidene difluoride (PVDF) binder in NMP (N-Methyl-2-Pyrrolidone) to form a slurry. The slurry was then uniformly casted onto carbon-coated Al foil. After drying, disk-shaped electrodes with a diameter of 15 mm were cut out from the casted layer and assembled in an Ar-filled glove box (UniLab, Braun, Germany) with lithium disk anodes (Ø 15.6 mm) and Celgard 2500 separators (Celgard LLC) in coin-type cells for electrochemical measurements, as shown in Scheme S1.[51,52] As electrolyte 20 μL mg$^{-1}$ of 1 M lithium bis(trifluoromethanesulfonyl) imide (LiTFSI) in 1, 3-dioxolane (DOL) and 1, 2-dimethoxymethane (DME) (1:1, v/v) with lithium nitrate (3 wt.%) was added in each cell. Cyclic voltammetry was performed at a scanning rate of 0.05 mV s$^{-1}$ on an electrochemical workstation (CHI 760e CH Instruments). The galvanostatic cyclic performance was measured at 0.2 C (1.0 C = 1670 mA g$^{-1}$) within a voltage range of 1.5 - 2.8 V. Before the cycling tests, the first discharge and charge of all cells was performed at 0.1 C for activation of the sulfur. Rate performances were measured at 0.1 C, 0.2 C, 0.5 C, 1.0 C, 2.0 C and back to 0.2 C.

**Material characterization**. X-ray diffraction (XRD) spectra were collected on a D8 Advance Bruker diffractometer with Cu K$_\alpha$ radiation employing a 0.25° divergent slit and a 0.125° anti-scattering slit; the patterns were recorded in the 2θ range from 5° to 80° with a 0.02° step and a counting time of 15 s per step.

Nitrogen adsorption-desorption isotherms were measured at -196 °C on a Micromeritics ASAP 2420 V2.05 porosimeter. Before the analysis, the samples were outgassed overnight at 120 °C under vacuum. The specific surface area was evaluated with the Brunauer-Emmett-Teller (BET) model by fitting the N$_2$ adsorption isotherm; the pore volume was determined at $P/P_0 = 0.995$ and the pore size distribution was obtained by applying the non-local density functional theory (NLDFT) method.

Thermogravimetric analysis (TGA) was performed using a Perkin Elmer Pyris Diamond TG/DTA. Samples of approximately 4 mg were heated in N$_2$ from 25 to 600 °C, at a rate of 5 °C min$^{-1}$.

Transmission electron microscope (TEM) images were obtained by using a FEI Tecnai T20, operating at 200 kV; the images were recorded under low-dose conditions with a slow scan CCD camera. A probe and image Cs aberration corrected 30-300 kV Thermo Fisher Scientific Thenis Z (scanning) transmission electron microscope (S/TEM) equipped with the dual X-ray detector was employed for the structural characterization of N-Ti$_3$C$_2$T$_x$-MIP. Images were acquired using high-angle annular dark-field (HAADF)-STEM (21 mrad convergence semi-angle, 50 pA probe current, 31-186 mrad collection angles of the HAADF detector), as well as bright-field and dark-field TEM. Scanning electronic microscopy (SEM) measurements were performed with a FEI Philips FEG-XL30s microscope; the morphology was characterized using a JEOL 2010 operating at 200 kV.

X-ray photoelectron spectroscopy (XPS) was performed with a Surface Science Instruments SSX-100 spectrometer, equipped with a monochromatic Al K$_\alpha$ X-ray source (hν =1486.6 eV). The measurement chamber pressure was maintained at $1\times10^{-9}$ mbar during data acquisition; the photoelectron take-off angle was 37° with respect to the surface normal. The diameter of the analyzed area was 1000 μm; the energy resolution was 1.26 eV (or 1.67 eV for a broad survey scan). Samples were dispersed in chloroform, sonicated and stirred for 30 min, and drop-casted on a thin gold film, grown on mica[53]. For the N-Ti$_3$C$_2$T$_x$-MIP/Li$_2$S$_8$, N-Ti$_3$C$_2$T$_x$-MIP was mixed with a Li$_2$S$_8$ solution and allowed to reach adsorption equilibrium after 24 hours of stirring. Unbonded Li$_2$S$_8$ was removed by repeat washing with DMF and centrifugation until the solution turned colorless, and the N-Ti$_3$C$_2$T$_x$-MIP/Li$_2$S$_8$ composite was drop-casted onto a gold-mica substrate; all steps were performed inside an Ar-filled glove box. Spectral analysis with the help of the least-squares curve-fitting program WinSpec (LISE, University of Namur, Belgium) included a Shirley background subtraction and fitting with a minimum number of peaks consistent with the expected composition of the probed volume, taking into account the experimental resolution. Peak profiles were taken as a convolution of Gaussian and Lorentzian functions; Binding energies (BEs) were referenced to Au $4f_{7/2}$ photoemission peak[54] centered at 84.0 eV and are accurate to ± 0.1 eV when deduced from the fitting procedure. All measurements were carried out on freshly prepared samples, and three different spots were measured on each sample to check for homogeneity.

## ASSOCIATED CONTENT

**Supporting Information**

The Supporting Information is available free of charge at *ACS Nano* website.

Schematic illustration of coin cell design, SEM images of exfoliated MXene sheets, XRD patterns of MAX phase and MXene, TEM image and FFT profile of $TiO_2$ on N-$Ti_3C_2T_x$-MIP, Wide scan XPS spectra of Ti3C2Tx, N-Ti3C2Tx, N-Ti3C2Tx-MIP, and N-Ti3C2Tx-MIP/S, and corresponding near-surface atomic percentage of each element. S2p core level region XPS of N-$Ti_3C_2T_x$-MIP/S.

## AUTHOR INFORMATION

**Corresponding Authors**

**Yutao Pei** - Engineering and Technology Institute Groningen, University of Groningen, Nijenborgh 4, 9747AG Groningen, the Netherlands. https://orcid.org/0000-0002-1817-2228;
Email: y.pei@rug.nl

**Petra Rudolf –** Zernike Institute for Advanced Materials, University of Groningen, Nijenborgh 4, 9747AG Groningen, the Netherlands. https://orcid.org/0000-0002-4418-1769; Email: p.rudolf@rug.nl

**Authors**

**Feng Yan -** Zernike Institute for Advanced Materials, University of Groningen, Nijenborgh 4, 9747AG Groningen, the Netherlands; http://orcid.org/0000-0002-9948-8450; current address: National Graphene Institute, University of Manchester, Manchester M13 9PL, United Kingdom

**Liqiang Li -** Engineering and Technology Institute Groningen, University of Groningen, Nijenborgh 4, 9747AG Groningen, the Netherlands; https://orcid.org/0000-0002-1531-0998; current address: Department for electrochemical energy storage, helmholtz-zentrum berlin für materialien und energie gmbh, Hahn-Meitner-Platz 1, 14109 Berlin

**Chongnan Ye -** Zernike Institute for Advanced Materials, University of Groningen, Nijenborgh 4, 9747AG Groningen, the Netherlands; https://orcid.org/0000-0003-2088-6913; current address: Institute of Chemical Sciences and Engineering (ISIC), École Polytechnique Fédérale de Lausanne (EPFL) Valais/Wallis, Sion, Switzerland

**Qi Chen -** Zernike Institute for Advanced Materials, University of Groningen, Nijenborgh 4, 9747AG Groningen, the Netherlands

**Sumit Kumar** - Electrical Engineering Division, Department of Engineering, University of Cambridge, Cambridge, CB21PZ, UK

**Wenjian Li** - Engineering and Technology Institute Groningen, University of Groningen, Nijenborgh 4, 9747AG Groningen, the Netherlands; https://orcid.org/0000-0002-9689-0823

**Hamoon Hemmatpour -** Zernike Institute for Advanced Materials, University of Groningen, Nijenborgh 4, 9747AG Groningen, the Netherlands; https://orcid.org/0000-0003-1858-203X;

**Konstantinos Spyrou -** Department of Materials Science and Engineering, University of Ioannina, 45110 Ioannina, Greece; https://orcid.org/0000-0002-2032-8439

**Sytze de Graaf -** Zernike Institute for Advanced Materials, University of Groningen, Nijenborgh 4, 9747AG Groningen, the Netherlands; https://orcid.org/0000-0002-0083-756X

**Marc C. A. Stuart -** Groningen Biomolecular Sciences and Biotechnology Institute, University of Groningen, Nijenborgh 4, 9747AG Groningen, the Netherlands; https://orcid.org/0000-0003-0667-6338

**Bart J. Kooi -** Zernike Institute for Advanced Materials, University of Groningen, Nijenborgh 4, 9747AG Groningen, the Netherlands; https://orcid.org/0000-0002-0311-4105

**Dimitrios P. Gournis -** Department of Materials Science and Engineering, University of Ioannina, 45110 Ioannina, Greece; https://orcid.org/0000-0003-4256-8190

**Katja Loos -** Zernike Institute for Advanced Materials, University of Groningen, Nijenborgh 4, 9747AG Groningen, the Netherlands; https://orcid.org/0000-0002-4613-1159

**Authors' contribution**

F.Y. and L.L. contributed equally to this work. F.Y. and L.L. initiated and P.R. and Y.P. supervised the project. F.Y. performed the material synthesis and wrote the manuscript with help of P.R.; L.L. carried out the battery performance characterization and data analysis with help from Y.P.; C.Y., Q.C., W.L. and K.L. participated in the experiments and theoretical discussion. S.K., K.S. and D.G. helped in the XPS data interpretation; M.C.A.S. helped in the TEM characterization. S. G. and B.J.K. helped in the STEM measurements. All authors discussed the results and commented on the manuscript.

**Conflict of Interest**

The authors declare no conflict of interest.

**ACKNOWLEDGEMENTS**

F.Y., Q.C. and W.L. gratefully thank for the financial support from China Scholarship Council. This work was supported by the Advanced Materials research program of the Zernike National Research Centre under the Bonus Incentive Scheme of the Dutch Ministry for Education, Culture and Science.

**TOC figure**

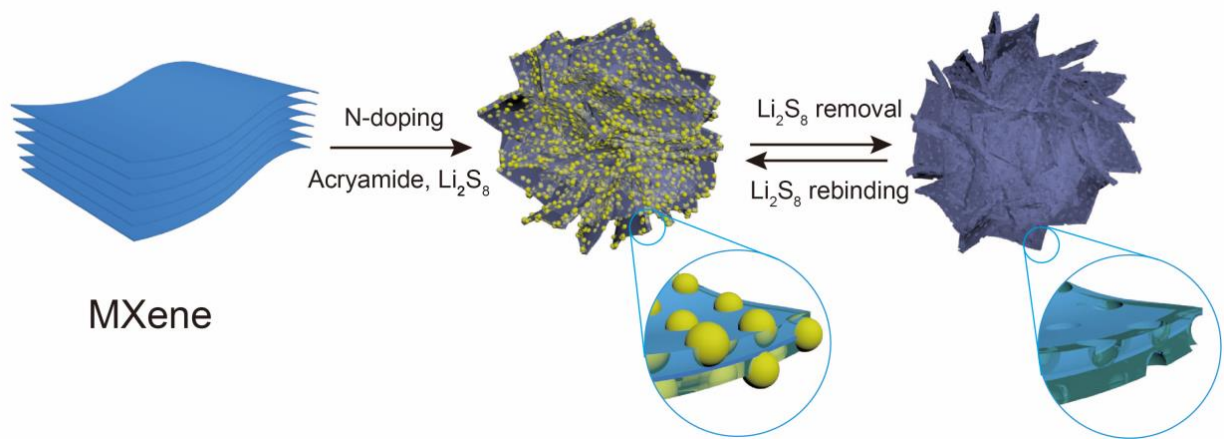